\documentclass[final,5p,times,twocolumn]{elsarticle}

\usepackage{graphicx}
\usepackage{amsmath}
\usepackage{amssymb}
\usepackage{hyperref}
\usepackage{latexsym}
\usepackage{epsfig}
\usepackage{epstopdf}
\usepackage{array}
\usepackage{ulem}
\usepackage[usenames,dvipsnames]{color}

\makeatletter
\def\@author#1{\g@addto@macro\elsauthors{\normalsize%
    \def\baselinestretch{1}%
    \upshape\authorsep#1\unskip\textsuperscript{%
      \ifx\@fnmark\@empty\else\unskip\sep\@fnmark\let\sep=,\fi
      \ifx\@corref\@empty\else\unskip\sep\@corref\let\sep=,\fi
      }%
    \def\authorsep{\unskip,\space}%
    \global\let\@fnmark\@empty
    \global\let\@corref\@empty  
    \global\let\sep\@empty}%
    \@eadauthor={#1}
}
\makeatother
\journal{Journal of Molecular Spectroscopy}

\begin{document}

\begin{frontmatter}

\title{State-specific detection of trapped HfF$^{+}$ by photodissociation}

\author{Kang-Kuen Ni\fnref{fn1}}
\fntext[fn1]{Present address: Department of Chemistry and Chemical Biology, Harvard University, Cambridge, MA 02138, USA}
\author{Huanqian Loh\fnref{fn2}}
\fntext[fn2]{Present address: Center for Quantum Technologies, National University of Singapore, Singapore 117543, Singapore}
\author{Matt Grau}
\author{Kevin C. Cossel\corref{cor2}}
\ead{cossel@jilau1.colorado.edu}
\author{Jun Ye }
\author{Eric A. Cornell\corref{cor2}}
\ead{cornell@jila.colorado.edu}

\cortext[cor2]{Corresponding Authors}

\address{JILA, National Institute of Standards and Technology and University of Colorado, and\\
Department of Physics University of Colorado, 440 UCB, Boulder, CO 80309, USA}

\date{\today}

\begin{abstract}
We use (1+1$'$) resonance-enhanced multiphoton photodissociation (REMPD) to detect the population in individual rovibronic states of trapped HfF$^+$ with a single-shot absolute efficiency of 18\%, which is over 200 times better than that obtained with fluorescence detection.  The first photon excites a specific rotational level to an intermediate vibronic band at 35,000--36,500~cm$^{-1}$, and the second photon, at 37,594~cm$^{-1}$ (266 nm), dissociates HfF$^+$ into Hf$^+$ and F. Mass-resolved time-of-flight ion detection then yields the number of state-selectively dissociated ions. Using this method, we observe rotational-state heating of trapped HfF$^+$ ions from collisions with neutral Ar atoms. Furthermore, we measure the lifetime of the $^3\Delta_1$ $v=0,\, J=1$ state to be 2.1(2)~s. This state will be used for a search for a permanent electric dipole moment of the electron.
\end{abstract}

\end{frontmatter}


\section{Introduction}

Quantum-state selective detection 
of molecular ions is a prerequisite for exciting new experiments in precision measurement~\cite{Leanhardt2011, Nguyen2011}, quantum information science~\cite{Schuster2011}, and astrochemistry~\cite{Snow2008}. 
Many proposed and ongoing precision measurement experiments utilize carefully chosen properties of molecules to improve previous constraints on fundamental symmetries and variations of constants set by experiments based on atoms \cite{Bressel2012, hudson2011, Bagdonaite2013, ACME2013}.
The hafnium fluoride ion (HfF$^+$) in its metastable ${^3}\Delta_{1}$ state, for example, can act as a sensitive probe for measurement of the electron's electric dipole moment (eEDM) due to its large effective internal electric field and many desirable properties for the evaluation of systematic uncertainties~\cite{Leanhardt2011, Loh2013}. Such a measurement relies on the precise determination of the molecular quantum state in the presence of both magnetic and electric fields. However, the detection of molecular states is often difficult and inefficient due to complex internal structure. Here we describe detection using resonance-enhanced multiphoton dissociation (REMPD) that overcomes these difficulties in HfF$^+$ and achieves a single-shot efficiency of 18(2)\%.


We use (1+1$'$) REMPD to break apart HfF$^+$ into Hf$^+$ and F and then count the Hf$^+$ products and remaining HfF$^+$ ions. The first photon in the (1+1$'$) scheme drives a bound-bound transition in HfF$^+$, making the photodissociation process rotational-state sensitive. Subsequently, the second photon excites HfF$^+$ from its intermediate state to a repulsive molecular potential. The two ion species are mass-resolved in time-of-flight to the detector. During our envisioned eEDM measurement~\cite{Loh2013}, the metastable $^3\Delta_1$ state will be populated via Raman transfer from the ground $X^1\Sigma^+$ electronic state. Therefore, it is desirable to find intermediate states that can be connected to both the $^3\Delta_1$ and $X^1\Sigma^+$ states using the first REMPD photon. In the subsequent sections, we report the HfF$^+$ photodissociation detection scheme (II) and the intermediate state spectrum in the 35,000--36,000 cm$^{-1}$ range (III).  Two applications utilize this new detection method: observation of rotational-state heating due to collisions (IV) and a measurement of the $^3\Delta_1$ metastable state lifetime (V).

\section{Photodissociation detection scheme}

Historically, a common molecular state read-out method is laser-induced fluorescence (LIF)~\cite{Kinsey1977}, 
where molecules in the target state are excited by a bound-to-bound transition and the subsequent spontaneously-emitted photons are measured. However, detecting fluorescence photons is often inefficient due to a small collection solid angle, low quantum efficiency of detectors at many wavelengths, and small branching ratios. The efficiency shortcoming of LIF can be alleviated if the molecules that occupy the quantum state of interest can be counted more directly, by taking advantage of high quantum efficiency ion detectors and easy manipulation of  ion trajectories so that the ion detectors effectively subtend $4\pi$ solid angle. Resonance-enhanced multiphoton ionization (REMPI), which allowed detection of individual neutral molecules~\cite{Grossman1977, Antonov1978}, has been utilized more recently for the efficient measurement of molecular quantum state populations. For detecting level population in molecular ions, the difficulty of efficiently and controllably removing the second electron makes REMPI a challenging technique to apply. 

An alternative to REMPI is to instead dissociate the molecular ion and monitor the dissociation products, leading to the development of REMPD~\cite{Weinkauf1987, Walter1988}. Soon after the first demonstrations REMPD was combined with ion trapping to perform spectroscopy on trapped molecular ions~\cite{Mikami1991}. Most applications relied on mass-selectivity to obtain zero-background spectra of excited molecular states, but have not used REMPD to sensitively measure lower state populations. Recent developments have focused on high-senstivity detection of small collections of trapped ions. In Roth et al.~\cite{Roth2006} and H{\o}jbjerre et al.~\cite{Hojbjerre2009}, rotational-state-resolved photodissociation was demonstrated with molecular ions in a Coulomb crystal. There, high signal-to-noise images of co-trapped atomic ions showed the photodissociated molecular ions as a loss of dark ions. While providing high quantum efficiency, this technique only works with Coulomb crystals. Rellergert et al.~\cite{Rellergert2013} instead used one-photon photodissociation followed by mass-resolved time-of-flight ion detection. This method was sensitive to the distribution of molecular vibrational states but did not provide rotational state selectivity. We demonstrate a combination of these techniques that provides high quantum efficiency, rotational state selectivity for multiple vibronic states, and rapid detection without the need for Coulomb crystals.

\begin{figure}[!t]
    \includegraphics[width=\columnwidth]{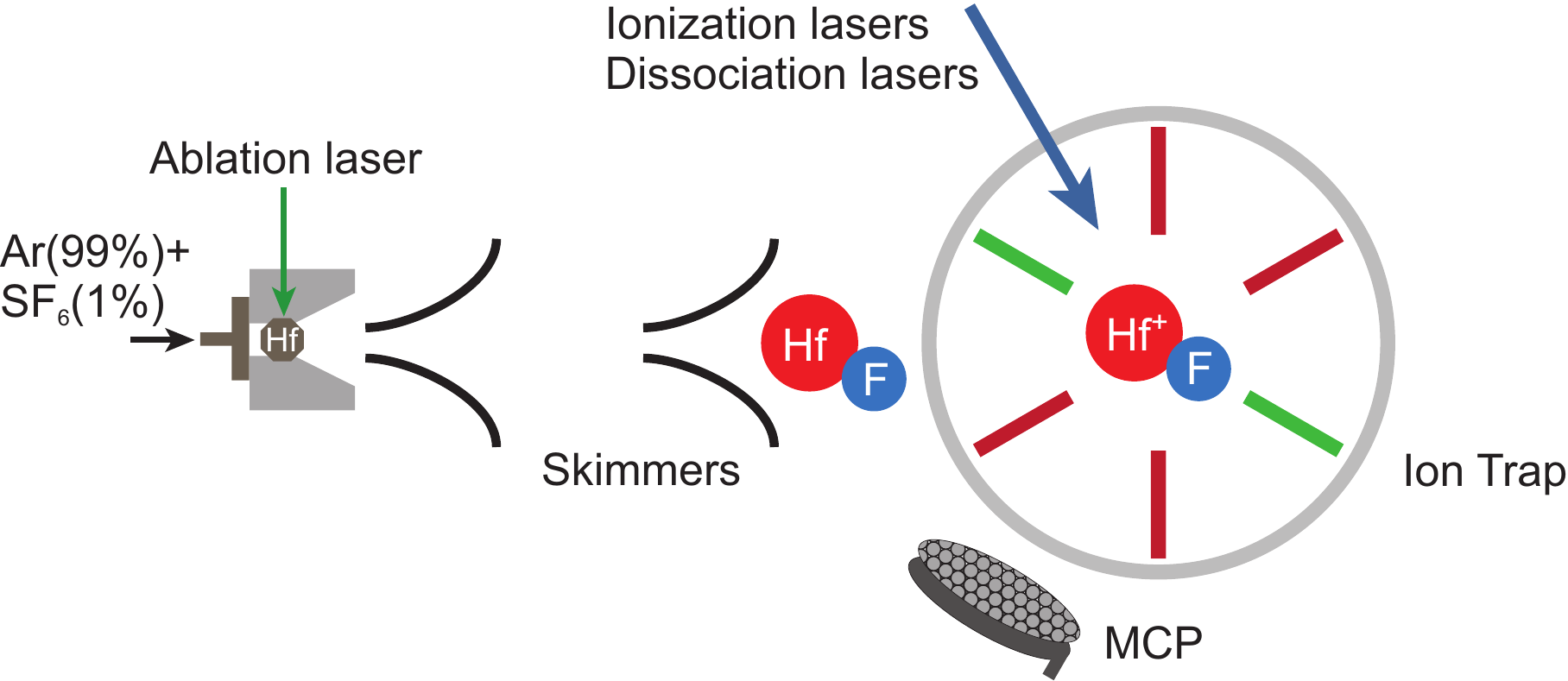}
  \caption{Schematic of the apparatus (top view). The apparatus consists of three regions: 1. HfF molecular source, generated by ablation of a Hf rod in the presence of SF$_6$ (1\%) + Argon (99\%) gas, 2. supersonic beam collimation through two skimmers, and 3. molecule ionization by lasers and trapping in a linear Paul trap with six radial trapping electrodes and two axial trapping end-caps (not shown). A microchannel plate (MCP), located 10~cm from the center of the trap, is used for time-of-flight ion detection.}
\label{app}
\end{figure}

To search for viable routes for photodissociation, we first prepared ions in a mixture of three different vibronic (vibrational and electronic) states: $X^1\Sigma^+(v=0)$, $X^1\Sigma^+(v=1)$  and $^3\Delta_1(v=0)$. The experimental sequence for ion preparation follows a similar scheme to that described in Ref.~\cite{loh2011}, and the apparatus is shown in Fig.~\ref{app}.
HfF neutral molecules are produced in a supersonic beam of 1\% of SF$_6$ and 99\% of Ar gas at 830~kPa backing pressure by ablation of a Hf rod just down-stream of a piezo valve. After two collimating skimmers, molecules in the beam enter the ion-trapping region where they are isotope-selectively ionized using two pulsed lasers (308~nm and 355~nm). Subsequently, the ion trap is turned on to confine the $^{180}$Hf$^{19}$F$^+$ ions. The ion trap is a specialized linear Paul trap with six radial confinement electrodes in order to apply a rotating bias field in addition to a quadrupole trapping potential. 

Prior to this work there was little information available regarding the locations of repulsive curves and intermediate states near 35,000 cm$^{-1}$. Any repulsive potentials must lie above the bond dissociation energy 51,200(1,600) cm$^{-1}$, which we determine by combining the measured neutral HfF bond dissociation energy \cite{Barkovskii1991} with the ionization energies of HfF \cite{Barker2011} and Hf \cite{Callender1988}. This estimate is in good agreement with that calculated from ab-initio theory \cite{Petrov2007}. Extrapolating from the calculated molecular potential of isoelectronic PtH$^+$~\cite{Meyer_note}, a $^3\Sigma$ repulsive potential of slope -80,000 cm$^{-1}/\AA$ is estimated to lie around 71,000 cm$^{-1}$ for HfF$^+$~\cite{Stutz}.

\begin{figure}[!t]
    \includegraphics[width=\columnwidth]{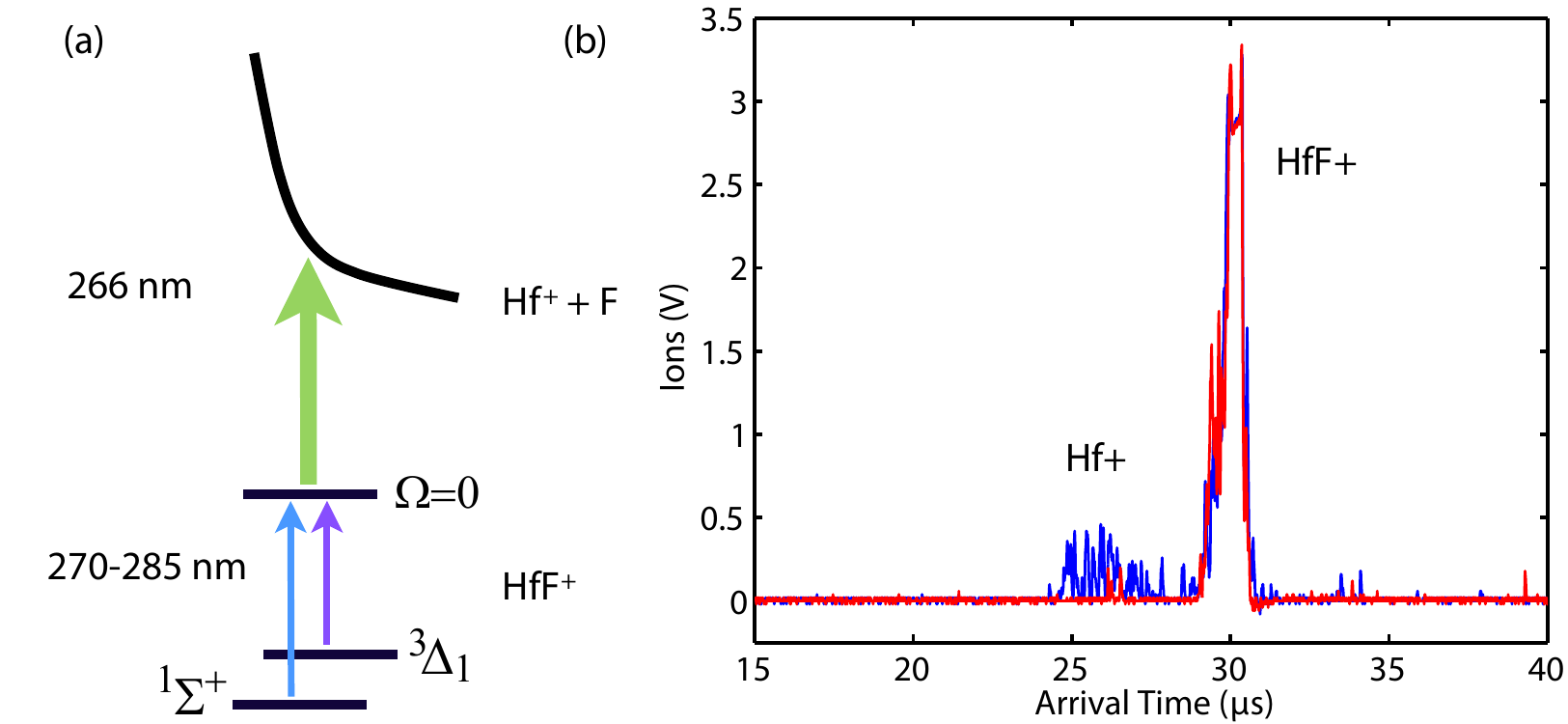} 
  \caption{(1+1$'$) REMPD detection. (a) Photodissociation pulses consist of two photons of different color. The first photon excites molecules from their initial state (one of  $X^1\Sigma^+$\, $v''\in\{0,1\}$ or  $^3\Delta_1$\, $v''=0$ to an intermediate state, and the second 266~nm photon dissociates the molecules into Hf$^+$ and F. (b) Mass-resolved time-of-flight ion detection signal, recorded without photodissociation (red) and with photodissociation (blue).  }
\label{scheme}
\end{figure}

To simplify the search for an unknown intermediate state energy and an unknown repulsive curve, we use a high power quadrupled Nd:YAG laser at 37,594~cm$^{-1}$ (266~nm, 5--10~mJ pulse energy) and a tunable laser in the range of 37,037--35,088 cm$^{-1}$ (270 to 285~nm, $< 1$~mJ pulse energy) for a total two-photon energy above 71,000 cm$^{-1}$ as shown in Fig~\ref{scheme}(a). Since the repulsive potential lies about 20,000~cm$^{-1}$ above the bond dissociation energy, the dissociated Hf$^+$ ions, being ten times heavier than dissociated F, can gain a near-isotropic root-mean-square velocity of 200~m/s. Velocity components transverse to the direction of travel to the MCP can compromise the effective solid angle subtended by the MCP: this is improved by applying a large impulse kick towards the MCP using the trap radial electrodes when the trap is turned off (Fig. \ref{app}). Further, the impulse kick is applied 100~$\mu$s after the photodissociation pulses, so that the ions' in-trap oscillations at 10~kHz can cause a ``refocusing'' of their motion along the transverse directions onto the MCP.

When no photodissociation pulse is applied, we see one large ion peak on the MCP centered 30 $\mu$s after the readout kick, which corresponds to the time-of-flight of HfF$^+$ (Fig.~\ref{scheme}(b)). If the photodissociation pulses are applied on resonance, a second, smaller number of Hf$^+$ ions appears at an earlier time. Prior to applying the dissociation lasers, the axial and radial confinement are ramped up to form a tighter trap. This helps to increase the dissociation efficiency, reduce the background counts, and also improve the mass resolution of the time-of-flight detection.

\section{Intermediate state spectrum}

\begin{figure}[!t]
	    \includegraphics[width=\columnwidth]{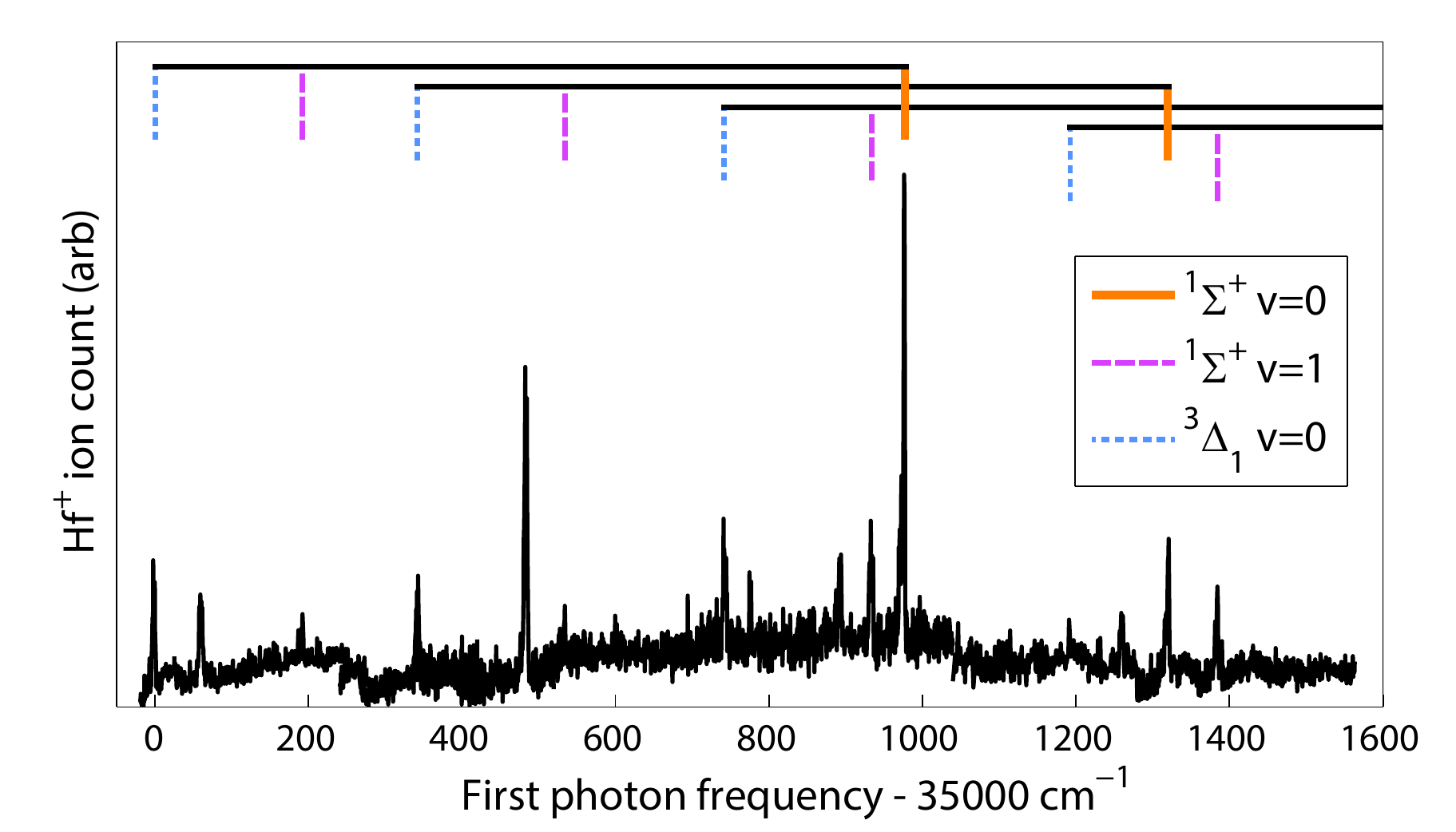} 
	  \caption{Photodissociation spectrum with the first photon tuned in the range of 35,000 to 36,500 cm$^{-1}$. We identify some lines originating from one of the three vibronic states:   $^3\Delta_1(v''=0)$, $X^1\Sigma^+(v''=1)$, and $X^1\Sigma^+(v''=0)$ by their respective spacings~\cite{Cossel2012}.}
	\label{spectrum}
\end{figure}

Figure~\ref{spectrum} shows a spectrum acquired by tuning the first photodissociation photon over 35,000--36,500 cm$^{-1}$ with the second photon fixed at 37,594 cm$^{-1}$ (266 nm). From the spectrum, we identify several sets of resonances (indicated by dashed and solid colored lines in Fig.~\ref{spectrum}) whose spacings correspond to the known vibronic spacings of the three states populated in the photoionization: $^3\Delta_1(v''=0)$, $X^1\Sigma^+(v''=1)$, and $X^1\Sigma^+(v''=0)$~\cite{Barker2011, Cossel2012}.  We further verify the assignments of $X^1\Sigma^+\,(v''=0)$ and $^3\Delta_1\,(v''=0)$ by performing depletion spectroscopy with a continuous-wave laser tuned to resonance with a particular transition. This laser optically pumps away ions in a given lower rotational state, which results in a decrease in the dissociation signal. In the case of $X^1\Sigma^+\,(v''=0)$, a depletion laser (770 nm) drives transitions to $^1\Pi_1\,(v'=0)$~\cite{loh2011}, whereas in the case of $^3\Delta_1\,(v''=0)$, a second depletion laser (684 nm) drives transitions to $^3\Phi_2\,(v'=1)$~\cite{Cossel2012}.

\begin{figure}[!t]
    \includegraphics[width=\columnwidth]{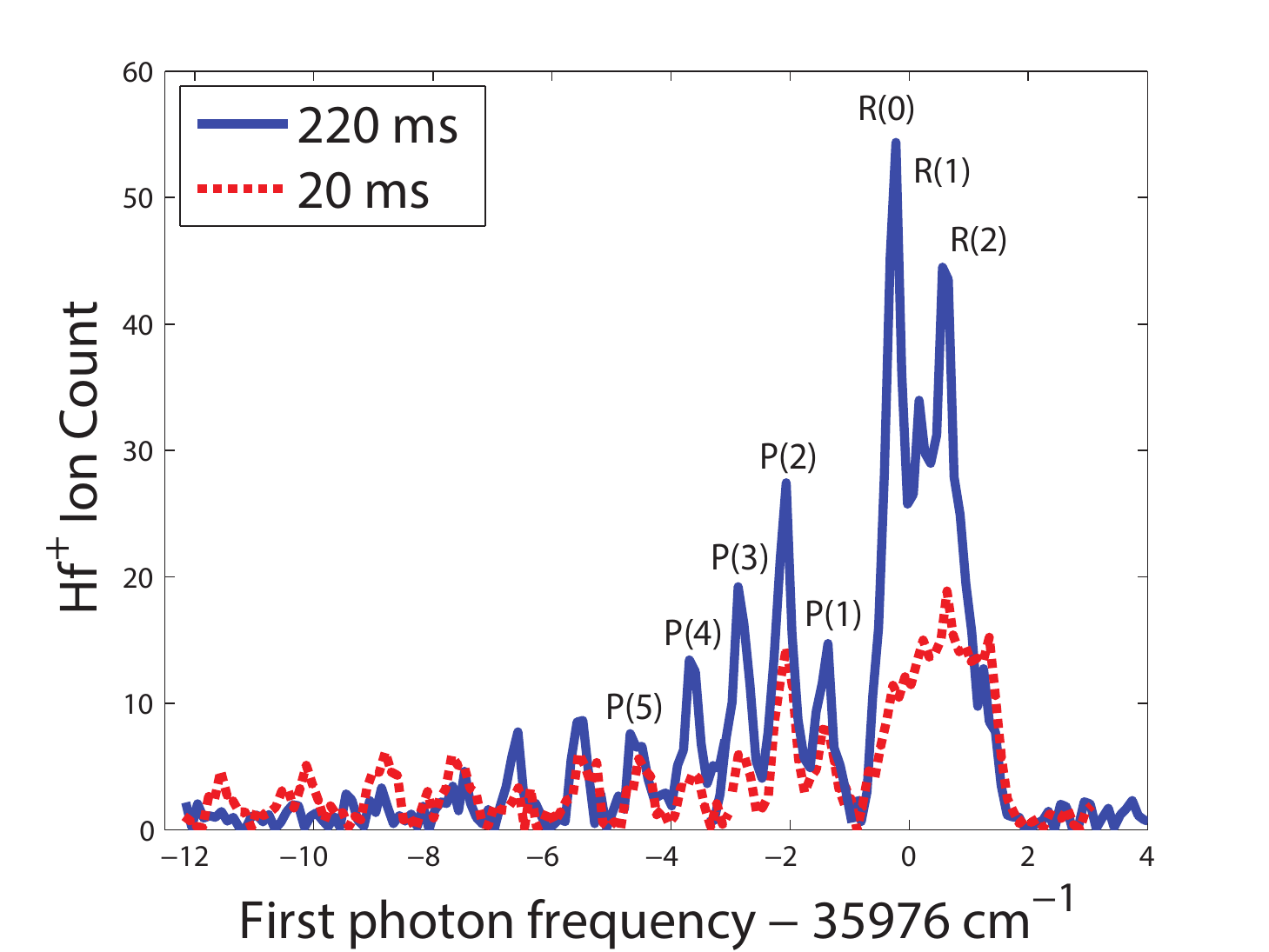} 
  \caption{Rotational state distribution of trapped ions in $X^1\Sigma^+\,v''=0$ state: 20~ms after trapping (blue) and 220~ms after trapping (red). The first REMPD photon is tuned near 35,976~cm$^{-1}$. We observed a redistribution towards higher rotational states after 220~ms of wait time.}
\label{rotdis}
\end{figure}

Our detection scheme is capable of resolving individual molecular rotational states. To record a detailed spectrum that shows the rotational distribution (see Fig.~\ref{rotdis}), we create molecular ions using the specific two-color autoionization transition (308~nm and 368~nm) described in \cite{loh2011}, where all the molecules are prepared in $X^1\Sigma^+\,(v''=0)$. For photodissociation detection, the pulsed dye laser that generates the first photon excitation has a linewidth of 0.12 cm$^{-1}$, a factor of six smaller than the spacings of $P$ and $R$ lines. The absence of $Q$ lines indicates that the intermediate state near 35,976 cm$^{-1}$ has a total angular-momentum-projection quantum number of $\Omega=0$. Furthermore, we fit the rotational constant of this intermediate state to be $B'=0.265(5)$~cm$^{-1}$.  


To calibrate the photodissociation detection efficiency, we probe the  $X^1\Sigma^+,\,v''=0,\,J''=0$ ions using the R(0) line of the 35,976 cm$^{-1}$ band shown in Fig.~\ref{rotdis}. About 35\% of the total ion population is initially in this state \cite{loh2011}. We observe that the photodissociation yield saturates with a laser fluence of 160(30)~$\mu$J/cm$^2$ for the first photon and approximately 50 mJ/cm$^2$ for the second photon. From the ratio of total HfF$^+$ counts on the MCP to the Hf$^+$ counts that are detected as products of state-sensitive photodissociation, we derive an efficiency of 32(2)\%. This includes a dissociation efficiency of 41(2)\%, which we measure from the fractional loss of HfF$^+$ signal when the (1+1$'$) pulse is applied, and a 78(2)\% transport efficiency of the dissociated ions to the detector.  The detector itself counts 57\% (90\% throughput from the grounded copper mesh in front of the detector, 63\% from the open area of the detector, and 100\% quantum efficiency for a given open channel) of the incident ions. Therefore the overall efficiency, including detector loss, is 18(2)\%.  Compared to our previous LIF detection \cite{loh2011}, this is an improvement by a factor of more than 200. Furthermore, the dissociation can be increased by applying multiple dissociation pulses (at a 10 Hz rate) and holding the dissociated Hf$^+$ in the trap between pulses. Population in the $^3\Delta_1\,v=0,\,J=1$ can be detected by dissociating through the same intermediate state using the 35,006 cm$^{-1}$ band. The yield on this transition saturates at a fluence of 70(10)~$\mu$J/cm$^2$.


In additional to rotationally resolved detection, hyperfine and parity quantum states of the molecules can be resolved by combining a narrow linewidth continuous wave laser for depletion with electric and magnetic fields to further separate the energy levels \cite{Loh2013}.

\section{Rotational heating}

Over the first 200~ms of trapping, we observe a redistribution towards higher rotational states compared to the initial rotational state distribution from the autoionization (see Fig.~\ref{rotdis}). We believe this state redistribution is due in part to Langevin collisions~\cite{Langevin1905} of HfF$^+$ ions with the neutral argon atoms that are injected as the supersonic beam's carrier gas. The gas is injected once per experimental cycle at a rate of 4 Hz, leading to a brief pressure spike when the ions are first loaded.  Figure~\ref{lifetime}(a) shows rapid initial loss from the $J = 0$ state (due to a relatively high argon pressure) followed by a much smaller loss rate after the argon has been pumped out of the vacuum chamber.  Further increasing the pumping speed may preserve a larger fraction of the ions that occupy the desired rotational state for subsequent experiments.

\section{Metastable state lifetime of $^3\Delta_1$}

\begin{figure}[!t]
    \includegraphics[width=\columnwidth]{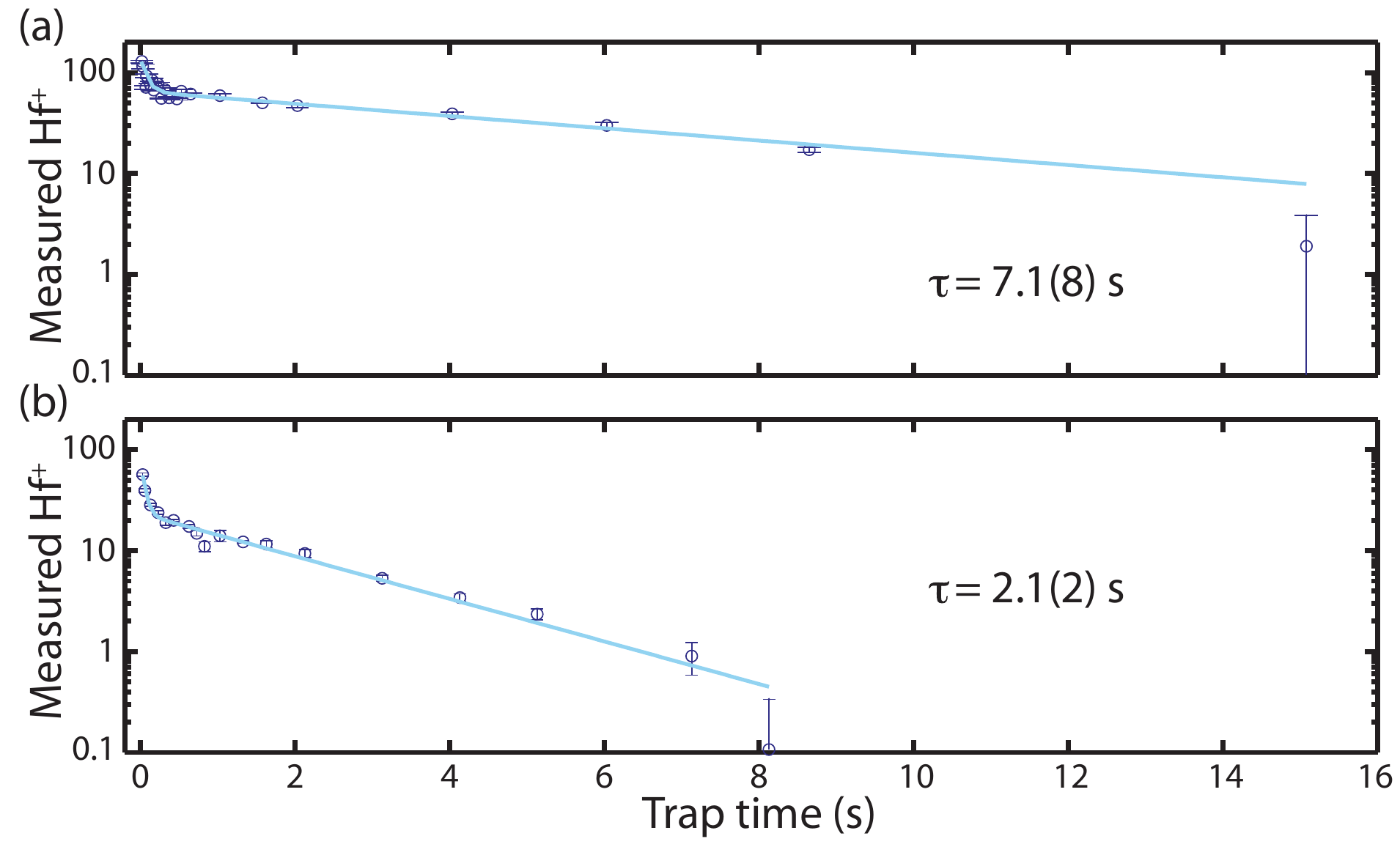} 
  \caption{Population lifetimes in trapped HfF$^+$. (a) $X^1\Sigma^+\,v=0,\,J=0$ and (b) $^3\Delta_1\,v=0,\,J=1$. The $X^1\Sigma^+$ lifetime exceeds many seconds after the initial fast decay (100 ms) due to rotational re-distribution as described in Section~IV. The $^3\Delta_1$ lifetime is 2.1(2)~s.
  }
\label{lifetime}
\end{figure}

Our detection capability allows measurement of the spontaneous-decay lifetime of the metastable $^3\Delta_1\, v=0,\, J=1$ state, which is of interest for eEDM searches~\cite{Leanhardt2011}. As shown in Fig.~\ref{lifetime}(b) the lifetime is 2.1(2)~s, which is in good agreement with the theoretical estimate of 1.7-2~s~\cite{Barker2011, Cossel2012}. For comparison, the lifetime of the ground $^1\Sigma^+\, v=0,\, J=0$ state is shown in Fig.~\ref{lifetime}(a). Blackbody excitation of rotational and vibrational levels is estimated to be 670~s and 6~s, respectively \cite{Stutz}. In neutral ThO, which was used in the most recent eEDM search \cite{ACME2013}, the radiative lifetime of the $^3\Delta_1$ state is $\sim1.8$~ms \cite{Vutha2010}. With HfF$^+$ having a lifetime three orders of magnitude longer, an eEDM measurement therein may realize much longer coherence times.

\section{Summary}

Ionic molecules are promising systems for a variety of studies, including precision measurement. However, quantum-state manipulation and detection are among the most challenging prerequisite for such studies. Here we demonstrate state-selective photodissociation combined with time-of-flight mass detection as a way of efficient quantum-state detection. Since HfF$^+$ consists of a heavy, fine-structure-active atom and a light, reactive atom, it has rich electronic structure and a deep potential well. We have located intermediate electronic states that connect to both (nominal) singlet and triplet electronic states. Using this detection scheme, we observe HfF$^+$ rotational heating within the first 200 ms of trapping. We also measure the metastable lifetime of the $^3\Delta{_1} \,v=0,\,J=1$ state to be 2.1(2)~s, which provides an upper limit on the coherence time available in an eEDM measurement. 

\section{Acknowledgments} 

This work is supported by the NSF, NIST and the Marsico Foundation. K.-K. Ni acknowledges a NIST/NRC Postdoctoral Fellowship. H. Loh was supported in part by A*STAR (Singapore). 

\bibliographystyle{apsrev}
\bibliography{refs}

\begin{thebibliography}{28}
\expandafter\ifx\csname natexlab\endcsname\relax\def\natexlab#1{#1}\fi
\expandafter\ifx\csname bibnamefont\endcsname\relax
  \def\bibnamefont#1{#1}\fi
\expandafter\ifx\csname bibfnamefont\endcsname\relax
  \def\bibfnamefont#1{#1}\fi
\expandafter\ifx\csname citenamefont\endcsname\relax
  \def\citenamefont#1{#1}\fi
\expandafter\ifx\csname url\endcsname\relax
  \def\url#1{\texttt{#1}}\fi
\expandafter\ifx\csname urlprefix\endcsname\relax\def\urlprefix{URL }\fi
\providecommand{\bibinfo}[2]{#2}
\providecommand{\eprint}[2][]{\url{#2}}

\bibitem[{\citenamefont{Leanhardt et~al.}(2011)\citenamefont{Leanhardt, Bohn,
  Loh, Maletinsky, Meyer, Sinclair, Stutz, and Cornell}}]{Leanhardt2011}
\bibinfo{author}{\bibfnamefont{A.~E.} \bibnamefont{Leanhardt}},
  \bibinfo{author}{\bibfnamefont{J.~L.} \bibnamefont{Bohn}},
  \bibinfo{author}{\bibfnamefont{H.}~\bibnamefont{Loh}},
  \bibinfo{author}{\bibfnamefont{P.}~\bibnamefont{Maletinsky}},
  \bibinfo{author}{\bibfnamefont{E.~R.} \bibnamefont{Meyer}},
  \bibinfo{author}{\bibfnamefont{L.~C.} \bibnamefont{Sinclair}},
  \bibinfo{author}{\bibfnamefont{R.~P.} \bibnamefont{Stutz}}, \bibnamefont{and}
  \bibinfo{author}{\bibfnamefont{E.~A.} \bibnamefont{Cornell}},
  \bibinfo{journal}{J. Mol. Spectrosc.} \textbf{\bibinfo{volume}{270}},
  \bibinfo{pages}{1} (\bibinfo{year}{2011}).

\bibitem[{\citenamefont{Nguyen and Odom}(2011)}]{Nguyen2011}
\bibinfo{author}{\bibfnamefont{J.~H.~V.} \bibnamefont{Nguyen}}
  \bibnamefont{and} \bibinfo{author}{\bibfnamefont{B.}~\bibnamefont{Odom}},
  \bibinfo{journal}{Physical Review A} \textbf{\bibinfo{volume}{83}},
  \bibinfo{pages}{053404} (\bibinfo{year}{2011}).

\bibitem[{\citenamefont{Schuster et~al.}(2011)\citenamefont{Schuster, Bishop,
  Chuang, DeMille, and Schoelkopf}}]{Schuster2011}
\bibinfo{author}{\bibfnamefont{D.~I.} \bibnamefont{Schuster}},
  \bibinfo{author}{\bibfnamefont{L.~S.} \bibnamefont{Bishop}},
  \bibinfo{author}{\bibfnamefont{I.~L.} \bibnamefont{Chuang}},
  \bibinfo{author}{\bibfnamefont{D.}~\bibnamefont{DeMille}}, \bibnamefont{and}
  \bibinfo{author}{\bibfnamefont{R.~J.} \bibnamefont{Schoelkopf}},
  \bibinfo{journal}{Phys. Rev. A} \textbf{\bibinfo{volume}{83}},
  \bibinfo{pages}{012311} (\bibinfo{year}{2011}).

\bibitem[{\citenamefont{Snow and Bierbaum}(2008)}]{Snow2008}
\bibinfo{author}{\bibfnamefont{T.~P.} \bibnamefont{Snow}} \bibnamefont{and}
  \bibinfo{author}{\bibfnamefont{V.~M.} \bibnamefont{Bierbaum}},
  \bibinfo{journal}{Annu. Rev. Anal. Chem.} \textbf{\bibinfo{volume}{1}},
  \bibinfo{pages}{229} (\bibinfo{year}{2008}).

\bibitem[{\citenamefont{Bressel et~al.}(2012)\citenamefont{Bressel, Borodin,
  Shen, Hansen, Ernsting, and Schiller}}]{Bressel2012}
\bibinfo{author}{\bibfnamefont{U.}~\bibnamefont{Bressel}},
  \bibinfo{author}{\bibfnamefont{A.}~\bibnamefont{Borodin}},
  \bibinfo{author}{\bibfnamefont{J.}~\bibnamefont{Shen}},
  \bibinfo{author}{\bibfnamefont{M.}~\bibnamefont{Hansen}},
  \bibinfo{author}{\bibfnamefont{I.}~\bibnamefont{Ernsting}}, \bibnamefont{and}
  \bibinfo{author}{\bibfnamefont{S.}~\bibnamefont{Schiller}},
  \bibinfo{journal}{Phys. Rev. Lett.} \textbf{\bibinfo{volume}{108}},
  \bibinfo{pages}{183003} (\bibinfo{year}{2012}).

\bibitem[{\citenamefont{Hudson et~al.}(2011)\citenamefont{Hudson, Kara,
  Smallman, Sauer, Tarbutt, and Hinds}}]{hudson2011}
\bibinfo{author}{\bibfnamefont{J.~J.} \bibnamefont{Hudson}},
  \bibinfo{author}{\bibfnamefont{D.~M.} \bibnamefont{Kara}},
  \bibinfo{author}{\bibfnamefont{I.~J.} \bibnamefont{Smallman}},
  \bibinfo{author}{\bibfnamefont{B.~E.} \bibnamefont{Sauer}},
  \bibinfo{author}{\bibfnamefont{M.~R.} \bibnamefont{Tarbutt}},
  \bibnamefont{and} \bibinfo{author}{\bibfnamefont{E.~A.} \bibnamefont{Hinds}},
  \bibinfo{journal}{Nature} \textbf{\bibinfo{volume}{473}},
  \bibinfo{pages}{493} (\bibinfo{year}{2011}).

\bibitem[{\citenamefont{Bagdonaite et~al.}(2013)\citenamefont{Bagdonaite,
  Jansen, Henkel, Bethlem, Menten, and Ubachs}}]{Bagdonaite2013}
\bibinfo{author}{\bibfnamefont{J.}~\bibnamefont{Bagdonaite}},
  \bibinfo{author}{\bibfnamefont{P.}~\bibnamefont{Jansen}},
  \bibinfo{author}{\bibfnamefont{C.}~\bibnamefont{Henkel}},
  \bibinfo{author}{\bibfnamefont{H.~L.} \bibnamefont{Bethlem}},
  \bibinfo{author}{\bibfnamefont{K.~M.} \bibnamefont{Menten}},
  \bibnamefont{and} \bibinfo{author}{\bibfnamefont{W.}~\bibnamefont{Ubachs}},
  \bibinfo{journal}{Science} \textbf{\bibinfo{volume}{339}},
  \bibinfo{pages}{46} (\bibinfo{year}{2013}).

\bibitem[{\citenamefont{{The ACME Collaboration}
  et~al.}(2013)\citenamefont{{The ACME Collaboration}, Baron, Campbell,
  DeMille, Doyle, Gabrielse, Gurevich, Hess, Hutzler, Kirilov
  et~al.}}]{ACME2013}
\bibinfo{author}{\bibnamefont{{The ACME Collaboration}}},
  \bibinfo{author}{\bibfnamefont{J.}~\bibnamefont{Baron}},
  \bibinfo{author}{\bibfnamefont{W.~C.} \bibnamefont{Campbell}},
  \bibinfo{author}{\bibfnamefont{D.}~\bibnamefont{DeMille}},
  \bibinfo{author}{\bibfnamefont{J.~M.} \bibnamefont{Doyle}},
  \bibinfo{author}{\bibfnamefont{G.}~\bibnamefont{Gabrielse}},
  \bibinfo{author}{\bibfnamefont{Y.~V.} \bibnamefont{Gurevich}},
  \bibinfo{author}{\bibfnamefont{P.~W.} \bibnamefont{Hess}},
  \bibinfo{author}{\bibfnamefont{N.~R.} \bibnamefont{Hutzler}},
  \bibinfo{author}{\bibfnamefont{E.}~\bibnamefont{Kirilov}},
  \bibnamefont{et~al.}, \bibinfo{journal}{Science}  (\bibinfo{year}{2013}),
  \eprint{http://www.sciencemag.org/content/early/2013/12/31/science.1248213.full.pdf}.

\bibitem[{\citenamefont{Loh et~al.}(2013)\citenamefont{Loh, Cossel, Grau, Ni,
  Meyer, Bohn, Ye, and Cornell}}]{Loh2013}
\bibinfo{author}{\bibfnamefont{H.}~\bibnamefont{Loh}},
  \bibinfo{author}{\bibfnamefont{K.}~\bibnamefont{Cossel}},
  \bibinfo{author}{\bibfnamefont{M.}~\bibnamefont{Grau}},
  \bibinfo{author}{\bibfnamefont{K.-K.} \bibnamefont{Ni}},
  \bibinfo{author}{\bibfnamefont{E.~R.} \bibnamefont{Meyer}},
  \bibinfo{author}{\bibfnamefont{J.~L.} \bibnamefont{Bohn}},
  \bibinfo{author}{\bibfnamefont{J.}~\bibnamefont{Ye}}, \bibnamefont{and}
  \bibinfo{author}{\bibfnamefont{E.~A.} \bibnamefont{Cornell}},
  \bibinfo{journal}{Science} \textbf{\bibinfo{volume}{342}},
  \bibinfo{pages}{1220} (\bibinfo{year}{2013}).

\bibitem[{\citenamefont{Kinsey}(1977)}]{Kinsey1977}
\bibinfo{author}{\bibfnamefont{J.}~\bibnamefont{Kinsey}},
  \bibinfo{journal}{Annu. Rev. Phys. Chem.} \textbf{\bibinfo{volume}{28}},
  \bibinfo{pages}{349} (\bibinfo{year}{1977}).

\bibitem[{\citenamefont{Grossman et~al.}(1977)\citenamefont{Grossman, Hurst,
  Payne, and Allman}}]{Grossman1977}
\bibinfo{author}{\bibfnamefont{L.}~\bibnamefont{Grossman}},
  \bibinfo{author}{\bibfnamefont{G.}~\bibnamefont{Hurst}},
  \bibinfo{author}{\bibfnamefont{M.}~\bibnamefont{Payne}}, \bibnamefont{and}
  \bibinfo{author}{\bibfnamefont{S.}~\bibnamefont{Allman}},
  \bibinfo{journal}{Chem. Phys. Lett.} \textbf{\bibinfo{volume}{50}},
  \bibinfo{pages}{70 } (\bibinfo{year}{1977}).

\bibitem[{\citenamefont{Antonov et~al.}(1978)\citenamefont{Antonov, Knyazev,
  Letokhov, Matiuk, Movshev, and Potapov}}]{Antonov1978}
\bibinfo{author}{\bibfnamefont{V.~S.} \bibnamefont{Antonov}},
  \bibinfo{author}{\bibfnamefont{I.~N.} \bibnamefont{Knyazev}},
  \bibinfo{author}{\bibfnamefont{V.~S.} \bibnamefont{Letokhov}},
  \bibinfo{author}{\bibfnamefont{V.~M.} \bibnamefont{Matiuk}},
  \bibinfo{author}{\bibfnamefont{V.~G.} \bibnamefont{Movshev}},
  \bibnamefont{and} \bibinfo{author}{\bibfnamefont{V.~K.}
  \bibnamefont{Potapov}}, \bibinfo{journal}{Opt. Lett.}
  \textbf{\bibinfo{volume}{3}}, \bibinfo{pages}{37} (\bibinfo{year}{1978}).

\bibitem[{\citenamefont{Weinkauf et~al.}(1987)\citenamefont{Weinkauf, Walter,
  Boesl, and Schlag}}]{Weinkauf1987}
\bibinfo{author}{\bibfnamefont{R.}~\bibnamefont{Weinkauf}},
  \bibinfo{author}{\bibfnamefont{K.}~\bibnamefont{Walter}},
  \bibinfo{author}{\bibfnamefont{U.}~\bibnamefont{Boesl}}, \bibnamefont{and}
  \bibinfo{author}{\bibfnamefont{E.}~\bibnamefont{Schlag}},
  \bibinfo{journal}{Chemical Physics Letters} \textbf{\bibinfo{volume}{141}},
  \bibinfo{pages}{267} (\bibinfo{year}{1987}).

\bibitem[{\citenamefont{Walter et~al.}(1988)\citenamefont{Walter, Weinkauf,
  Boesl, and Schlag}}]{Walter1988}
\bibinfo{author}{\bibfnamefont{K.}~\bibnamefont{Walter}},
  \bibinfo{author}{\bibfnamefont{R.}~\bibnamefont{Weinkauf}},
  \bibinfo{author}{\bibfnamefont{U.}~\bibnamefont{Boesl}}, \bibnamefont{and}
  \bibinfo{author}{\bibfnamefont{E.~W.} \bibnamefont{Schlag}},
  \bibinfo{journal}{The Journal of Chemical Physics}
  \textbf{\bibinfo{volume}{89}}, \bibinfo{pages}{1914} (\bibinfo{year}{1988}).

\bibitem[{\citenamefont{Mikami et~al.}(1991)\citenamefont{Mikami, Sasaki, and
  Sato}}]{Mikami1991}
\bibinfo{author}{\bibfnamefont{N.}~\bibnamefont{Mikami}},
  \bibinfo{author}{\bibfnamefont{T.}~\bibnamefont{Sasaki}}, \bibnamefont{and}
  \bibinfo{author}{\bibfnamefont{S.}~\bibnamefont{Sato}},
  \bibinfo{journal}{Chemical Physics Letters} \textbf{\bibinfo{volume}{180}},
  \bibinfo{pages}{431} (\bibinfo{year}{1991}).

\bibitem[{\citenamefont{Roth et~al.}(2006)\citenamefont{Roth, Koelemeij, Daerr,
  and Schiller}}]{Roth2006}
\bibinfo{author}{\bibfnamefont{B.}~\bibnamefont{Roth}},
  \bibinfo{author}{\bibfnamefont{J.~C.~J.} \bibnamefont{Koelemeij}},
  \bibinfo{author}{\bibfnamefont{H.}~\bibnamefont{Daerr}}, \bibnamefont{and}
  \bibinfo{author}{\bibfnamefont{S.}~\bibnamefont{Schiller}},
  \bibinfo{journal}{Phys. Rev. A} \textbf{\bibinfo{volume}{74}},
  \bibinfo{pages}{040501} (\bibinfo{year}{2006}).

\bibitem[{\citenamefont{H{\o}jbjerre et~al.}(2009)\citenamefont{H{\o}jbjerre,
  Hansen, Skyt, Staanum, and Drewsen}}]{Hojbjerre2009}
\bibinfo{author}{\bibfnamefont{K.}~\bibnamefont{H{\o}jbjerre}},
  \bibinfo{author}{\bibfnamefont{A.~K.} \bibnamefont{Hansen}},
  \bibinfo{author}{\bibfnamefont{P.~S.} \bibnamefont{Skyt}},
  \bibinfo{author}{\bibfnamefont{P.~F.} \bibnamefont{Staanum}},
  \bibnamefont{and} \bibinfo{author}{\bibfnamefont{M.}~\bibnamefont{Drewsen}},
  \bibinfo{journal}{New J. Phys.} \textbf{\bibinfo{volume}{11}},
  \bibinfo{pages}{055026} (\bibinfo{year}{2009}).

\bibitem[{\citenamefont{Rellergert et~al.}(2013)\citenamefont{Rellergert,
  Sullivan, Schowalter, Kotochigova, Chen, and Hudson}}]{Rellergert2013}
\bibinfo{author}{\bibfnamefont{W.~G.} \bibnamefont{Rellergert}},
  \bibinfo{author}{\bibfnamefont{S.~T.} \bibnamefont{Sullivan}},
  \bibinfo{author}{\bibfnamefont{S.~J.} \bibnamefont{Schowalter}},
  \bibinfo{author}{\bibfnamefont{S.}~\bibnamefont{Kotochigova}},
  \bibinfo{author}{\bibfnamefont{K.}~\bibnamefont{Chen}}, \bibnamefont{and}
  \bibinfo{author}{\bibfnamefont{E.~R.} \bibnamefont{Hudson}},
  \bibinfo{journal}{Nature} \textbf{\bibinfo{volume}{495}},
  \bibinfo{pages}{490} (\bibinfo{year}{2013}).

\bibitem[{\citenamefont{Loh et~al.}(2011)\citenamefont{Loh, Wang, Grau, Yahn,
  Field, Greene, and Cornell}}]{loh2011}
\bibinfo{author}{\bibfnamefont{H.}~\bibnamefont{Loh}},
  \bibinfo{author}{\bibfnamefont{J.}~\bibnamefont{Wang}},
  \bibinfo{author}{\bibfnamefont{M.}~\bibnamefont{Grau}},
  \bibinfo{author}{\bibfnamefont{T.~S.} \bibnamefont{Yahn}},
  \bibinfo{author}{\bibfnamefont{R.~W.} \bibnamefont{Field}},
  \bibinfo{author}{\bibfnamefont{C.~H.} \bibnamefont{Greene}},
  \bibnamefont{and} \bibinfo{author}{\bibfnamefont{E.~A.}
  \bibnamefont{Cornell}}, \bibinfo{journal}{J. Chem. Phys.}
  \textbf{\bibinfo{volume}{135}}, \bibinfo{pages}{154308}
  (\bibinfo{year}{2011}).

\bibitem[{\citenamefont{Barkovskii et~al.}(1991)\citenamefont{Barkovskii,
  Tsirel'nikov, Emel'yanov, and Khodeev}}]{Barkovskii1991}
\bibinfo{author}{\bibfnamefont{N.~V.} \bibnamefont{Barkovskii}},
  \bibinfo{author}{\bibfnamefont{V.~I.} \bibnamefont{Tsirel'nikov}},
  \bibinfo{author}{\bibfnamefont{A.~M.} \bibnamefont{Emel'yanov}},
  \bibnamefont{and} \bibinfo{author}{\bibfnamefont{Y.~S.}
  \bibnamefont{Khodeev}}, \bibinfo{journal}{Teplofiz.\ Vysok.\ Temper.}
  \textbf{\bibinfo{volume}{29}}, \bibinfo{pages}{474} (\bibinfo{year}{1991}).

\bibitem[{\citenamefont{Barker et~al.}(2011)\citenamefont{Barker, Antonov,
  Bondybey, and Heaven}}]{Barker2011}
\bibinfo{author}{\bibfnamefont{B.~J.} \bibnamefont{Barker}},
  \bibinfo{author}{\bibfnamefont{I.~O.} \bibnamefont{Antonov}},
  \bibinfo{author}{\bibfnamefont{V.~E.} \bibnamefont{Bondybey}},
  \bibnamefont{and} \bibinfo{author}{\bibfnamefont{M.~C.}
  \bibnamefont{Heaven}}, \bibinfo{journal}{J.\ Chem.\ Phys.}
  \textbf{\bibinfo{volume}{134}}, \bibinfo{pages}{201102}
  (\bibinfo{year}{2011}).

\bibitem[{\citenamefont{Callender et~al.}(1988)\citenamefont{Callender,
  Hackett, and Rayner}}]{Callender1988}
\bibinfo{author}{\bibfnamefont{C.~L.} \bibnamefont{Callender}},
  \bibinfo{author}{\bibfnamefont{P.~A.} \bibnamefont{Hackett}},
  \bibnamefont{and} \bibinfo{author}{\bibfnamefont{D.~M.}
  \bibnamefont{Rayner}}, \bibinfo{journal}{J.\ Opt.\ Soc.\ Am.\ B}
  \textbf{\bibinfo{volume}{5}}, \bibinfo{pages}{1341} (\bibinfo{year}{1988}).

\bibitem[{\citenamefont{Petrov et~al.}(2007)\citenamefont{Petrov, Mosyagin,
  Isaev, and Titov}}]{Petrov2007}
\bibinfo{author}{\bibfnamefont{A.~N.} \bibnamefont{Petrov}},
  \bibinfo{author}{\bibfnamefont{N.~S.} \bibnamefont{Mosyagin}},
  \bibinfo{author}{\bibfnamefont{T.~A.} \bibnamefont{Isaev}}, \bibnamefont{and}
  \bibinfo{author}{\bibfnamefont{A.~V.} \bibnamefont{Titov}},
  \bibinfo{journal}{Phys. Rev. A} \textbf{\bibinfo{volume}{76}},
  \bibinfo{pages}{030501(R)} (\bibinfo{year}{2007}).

\bibitem[{\citenamefont{Meyer and Bohn}(2012)}]{Meyer_note}
\bibinfo{author}{\bibfnamefont{E.~R.} \bibnamefont{Meyer}} \bibnamefont{and}
  \bibinfo{author}{\bibfnamefont{J.~L.} \bibnamefont{Bohn}}
  (\bibinfo{year}{2012}), \bibinfo{note}{{Private communication}}.

\bibitem[{\citenamefont{Stutz}({2010})}]{Stutz}
\bibinfo{author}{\bibfnamefont{R.~P.} \bibnamefont{Stutz}}, Ph.D. thesis,
  \bibinfo{school}{University of Colorado - Boulder} (\bibinfo{year}{{2010}}).

\bibitem[{\citenamefont{Cossel et~al.}(2012)\citenamefont{Cossel, Gresh,
  Sinclair, Coffey, Skripnikov, Petrov, Mosyagin, Titov, Field, Meyer
  et~al.}}]{Cossel2012}
\bibinfo{author}{\bibfnamefont{K.~C.} \bibnamefont{Cossel}},
  \bibinfo{author}{\bibfnamefont{D.~N.} \bibnamefont{Gresh}},
  \bibinfo{author}{\bibfnamefont{L.~C.} \bibnamefont{Sinclair}},
  \bibinfo{author}{\bibfnamefont{T.}~\bibnamefont{Coffey}},
  \bibinfo{author}{\bibfnamefont{L.~V.} \bibnamefont{Skripnikov}},
  \bibinfo{author}{\bibfnamefont{A.~N.} \bibnamefont{Petrov}},
  \bibinfo{author}{\bibfnamefont{N.~S.} \bibnamefont{Mosyagin}},
  \bibinfo{author}{\bibfnamefont{A.~V.} \bibnamefont{Titov}},
  \bibinfo{author}{\bibfnamefont{R.~W.} \bibnamefont{Field}},
  \bibinfo{author}{\bibfnamefont{E.~R.} \bibnamefont{Meyer}},
  \bibnamefont{et~al.}, \bibinfo{journal}{Chem. Phys. Lett.}
  \textbf{\bibinfo{volume}{546}}, \bibinfo{pages}{1 } (\bibinfo{year}{2012}).

\bibitem[{\citenamefont{Langevin}(1905)}]{Langevin1905}
\bibinfo{author}{\bibfnamefont{M.~P.} \bibnamefont{Langevin}},
  \bibinfo{journal}{Ann. Chim. Phys.} \textbf{\bibinfo{volume}{5}},
  \bibinfo{pages}{245} (\bibinfo{year}{1905}).

\bibitem[{\citenamefont{Vutha et~al.}(2010)\citenamefont{Vutha, Campbell,
  Gurevich, Hutzler, Parsons, Patterson, Petrik, Spaun, Doyle, Gabrielse
  et~al.}}]{Vutha2010}
\bibinfo{author}{\bibfnamefont{A.~C.} \bibnamefont{Vutha}},
  \bibinfo{author}{\bibfnamefont{W.~C.} \bibnamefont{Campbell}},
  \bibinfo{author}{\bibfnamefont{Y.~V.} \bibnamefont{Gurevich}},
  \bibinfo{author}{\bibfnamefont{N.~R.} \bibnamefont{Hutzler}},
  \bibinfo{author}{\bibfnamefont{M.}~\bibnamefont{Parsons}},
  \bibinfo{author}{\bibfnamefont{D.}~\bibnamefont{Patterson}},
  \bibinfo{author}{\bibfnamefont{E.}~\bibnamefont{Petrik}},
  \bibinfo{author}{\bibfnamefont{B.}~\bibnamefont{Spaun}},
  \bibinfo{author}{\bibfnamefont{J.~M.} \bibnamefont{Doyle}},
  \bibinfo{author}{\bibfnamefont{G.}~\bibnamefont{Gabrielse}},
  \bibnamefont{et~al.}, \bibinfo{journal}{J. Phys. B}
  \textbf{\bibinfo{volume}{43}}, \bibinfo{pages}{074007}
  (\bibinfo{year}{2010}).

\end{thebibliography}

\end{document}